\documentclass[aps,prl,twocolumn,showpacs,amsmath,amssymb]{revtex4}
\usepackage[latin2]{inputenc}
\usepackage{amsmath}
\usepackage{graphicx}
\usepackage{multirow}
\usepackage{dcolumn}
\usepackage{epstopdf}

\newcommand{\ba}{\begin{eqnarray*}}
\newcommand{\ea}{\end{eqnarray*}}
\newcommand{\baa}{\begin{eqnarray}}
\newcommand{\eaa}{\end{eqnarray}}
\def\bar{\begin{array}}
\def\ear{\end{array}}
\def\LB{\left(}
\def\RB{\right)}

\def\s{\sigma}
\def\f{\frac}

\def\nn{\nonumber}
\def\R{\mathbf{R}}

\begin{document}

\title{Exact factorization-based density functional theory of electrons and nuclei} 

\author{Ryan Requist}
\email{rrequist@mpi-halle.mpg.de}
\affiliation{
Max Planck Institute of Microstructure Physics, Weinberg 2, 06120 Halle, Germany 
}
\author{E. K. U. Gross}
\affiliation{
Max Planck Institute of Microstructure Physics, Weinberg 2, 06120 Halle, Germany
}

\date{\today}

\begin{abstract}
The ground state energy of a system of electrons and nuclei is proven to be a variational functional of the conditional electronic density $n_R(\mathbf{r})$, the nuclear wavefunction $\chi(R)$ and an induced vector potential $A_{\mu}(R)$ and quantum geometric tensor $\mathcal{T}_{\mu\nu}(R)$ derived from the conditional electronic wavefunction $\Phi_R(r)$ over nuclear configuration space, where $r=\mathbf{r}_1,\mathbf{r}_2,\ldots$ are electronic coordinates and $R=\R_1,\R_2,\ldots$ are nuclear coordinates.  The ground state $(n_R,\chi,A_{\mu},\mathcal{T}_{\mu\nu})$ can be calculated by solving self-consistently (i) conditional Kohn-Sham equations containing an effective potential $v_{\rm s}(\mathbf{r})$ that depends parametrically on $R$, (ii) the Schr\"odinger equation for $\chi(R)$ and (iii) Euler-Lagrange equations that determine $\mathcal{T}_{\mu\nu}$. The theory is applied to the $E\otimes e$ Jahn-Teller model. 
\end{abstract}


\maketitle

The foundations of density functional theory (DFT) \cite{hohenberg1964,kohn1965} are inextricably tied to the Born-Oppenheimer approximation.  In DFT applications, e.g.~electronic band structure calculations, it often suffices to treat the nuclei classically or to fix them to their equilibrium positions.  Quantum nuclear effects such as tunneling, delocalization and zero-point energy are, however, relevant for several interesting problems, e.g.~the phases of ice \cite{stillinger1983,lee1992,roettger2012,pamuk2012} and the local structure of water \cite{soper2008,habershon2009,li2011,ceriotti2016}, and were recently reported to enable thermally-activated tunneling of protons through a graphene layer \cite{lozada-hidalgo2016,poltavsky2016}.  Some quantum nuclear effects can be included in DFT-based calculations by quantizing nuclear vibrations on the adiabatic ground state potential energy surface, but because such an approach relies on the Born-Oppenheimer approximation, it is not formally exact.  When the nuclear variables and electron-nuclear coupling are treated exactly and fully quantum mechanically, the electrons feel, instead of the external potential $v(\mathbf{r})$ of DFT, a ``weighted'' potential $-\sum_i \int |\chi(R)|^2 Z_i e^2/|\mathbf{r}-\mathbf{R}_i| dR$, modified by the delocalization of the nuclear probability density $|\chi(R)|^2$, but also additional interactions induced by nonadiabatic electron-nuclear correlations \cite{gidopoulos2014,abedi2010} not included in standard DFT functionals.

Particularly in time-dependent processes such as photoinduced chemical bond dynamics \cite{zewail2000}, proton transfer in hydrogen-bonded systems \cite{marx2010}, dissociative adsorption of H$_2$ on Pd(100) \cite{gross2000}, and molecular processes involving conical intersections of Born-Oppenheimer (BO) potential energy surfaces \cite{domcke2011}, nonadiabatic and quantum nuclear effects may be significant.  Mixed quantum-classical approaches, which couple quantum mechanical electrons to classical nuclear motion, usually adopt an effective single-particle description of the electrons, and DFT is often the only method capable of treating large systems of interest with sufficient accuracy.

For the further development of theories capable of describing quantum nuclear effects in large systems, it would be useful to know whether it is in principle possible to include full quantum nuclear motion and electronic-vibrational coupling while retaining a density-functional formulation of the electronic part of the problem.  One way to answer this question is to define a multicomponent DFT in terms of an electronic density $\rho(\mathbf{r})$ in the body-fixed frame of the nuclei and an $N_n$-body nuclear density $\Gamma(R)=\int |\Psi(r,R)|^2 dr$, where $N_n$ is the number of nuclei.  A Hohenberg-Kohn-type theorem establishing a one-to-one correspondence between the densities $\{\rho(\mathbf{r}),\Gamma(R)\}$ and auxiliary potentials $\{v(\mathbf{r}),V(R)\}$ has been proven \cite{kreibich2001,gidopoulos1998,butriy2007}.  To use this theory, one needs an approximation for a Hartree-exchange-correlation functional $E_{\rm hxc}[\rho,\Gamma]$ depending on both densities.  

Here, we pursue a different approach that is also exact in principle and allows one to reuse the well-developed exchange-correlation functionals of DFT at the first level of approximation.  Being built on the exact factorization scheme \cite{hunter1975,gidopoulos2014,abedi2010}, our approach incorporates the true nuclear Schr\"odinger equation, including induced scalar and vector potentials.  The main objective of this Letter is to prove that the ground state energy is a variational functional of (i) the conditional electronic density $n_R(\mathbf{r})$, (ii) the nuclear wavefunction $\chi(R)$ and (iii) an induced vector potential $A_{\mu}(R)$ and quantum geometric tensor $\mathcal{T}_{\mu\nu}(R)$ responsible for electromagnetic effects in the nuclear Schr\"odinger equation \cite{berry1989,berry1990,requist2016}. We propose a minimization scheme that preserves the single-particle picture of DFT while including full quantum nuclear effects and electronic-vibrational coupling. 

\noindent \textbf{\textit{Exact factorization.}} 
We start from the nonrelativistic Hamiltonian of a system of $N_{\rm e}$ electrons and $N_{\rm n}$ nuclei,
\begin{align}
\hat{H} = -\sum_{i=1}^{N_{\rm n}} \f{\hbar^2\nabla_{\mathbf{R}_i}^2}{2M_i} -\sum_{i=1}^{N_{\rm e}} \f{\hbar^2\nabla_{\mathbf{r}_i}^2}{2m_{\rm e}} + \hat{V}_{\rm nn} + \hat{V}_{\rm ee} + \hat{V}_{\rm en} {,}
\label{eq:H}
\end{align}
where the Coulomb interaction between nuclei is
\begin{align}
\hat{V}_{\rm nn} = \f{1}{4\pi\varepsilon_0} \sum_i \sum_{j<i} \f{Z_i Z_j e^2}{|\mathbf{R}_i-\mathbf{R}_j|} {,}
\end{align}
and the electron-electron ($\hat{V}_{\rm ee}$) and electron-nucleus ($\hat{V}_{\rm en}$) terms are defined analogously. It was shown \cite{hunter1975} that the full wavefunction can be written exactly in the factorized form $\Psi(r,R)=\Phi_R(r) \chi(R)$, where $\Phi_R(r)$ is a conditional electronic wavefunction depending parametrically on the nuclear coordinates and obeying the partial normalization condition $\int |\Phi_R(r)|^2 dr=1$ for all $R$ and $\chi(R)$ is the marginal nuclear wavefunction.  $\Phi_R(r)$ and $\chi(R)$ are determined by a pair of coupled equations \cite{gidopoulos2014,abedi2010}.  

Since the Hamiltonian is translationally invariant, any eigenstate $\Psi(r,R)$ belongs to a continuum.  To get square integrable eigenstates, we change coordinates from $(r,R)$ to $(q,Q,\textbf{R}_{\rm cm})$, where $\mathbf{R}_{\rm cm}$ is the total center of mass, $Q=(Q^1,Q^2,\ldots)$ represents a set of $3(N-1)$ generalized nuclear coordinates and $q$ is the set of electronic coordinates referred to the nuclear center of mass \cite{sutcliffe2000}.  After dividing off a function of $\mathbf{R}_{\rm cm}$, the exact factorization scheme can be used to write the remaining wavefunction as $\Phi_Q(q) \chi(Q)$ and to derive a pair of coupled equations for the factors $\Phi_Q(q)$ and $\chi(Q)$ that are formally equivalent to those for $\Phi_R(r)$ and $\chi(R)$ in Ref.~\cite{gidopoulos2014}, except for modifications to the Hamiltonian operators (see Supplemental Material).  Thus, we will change our notations and from now on let $R$ denote the set $Q$ and $r$ the set $q$.  The equations for the factors $\Phi_R(r)$ and $\chi(R)$ are then \begin{align}
&\Big[ \hat{H}^{BO} + \f{1}{2}  \mathcal{I}^{\mu\nu} (P_{\mu}-A_{\mu})(P_{\nu}-A_{\nu}) \nn \\
&\quad+  \mathcal{I}^{\mu\nu} \Big( \f{P_{\mu} \chi}{\chi} + A_{\mu} \Big) ( P_{\nu}-A_{\nu}) \Big] |\Phi_R\rangle = \mathcal{E}(R) |\Phi_R\rangle {,} \label{eq:Phi} \\[0.2cm]
&\Big[ \f{1}{2} \mathcal{I}^{\mu\nu} (P_{\mu} + A_{\mu})(P_{\nu} + A_{\nu}) +  \mathcal{E}(R) \Big] \chi(R) = E \chi(R) {,} \label{eq:chi}
\end{align}
where $\hat{H}^{\rm BO} =\hat{H} - \hat{T}_n - \f{\hat{\mathbf{P}}_{\rm cm}^2}{2M_{\rm tot}}$ and the nuclear kinetic energy operator has been put in the Watson-type form $\hat{T}_{\rm n} = \f{1}{2} \mathcal{I}^{\mu\nu} P_{\mu} P_{\nu}$ \cite{watson1968} with an inverse inertia tensor $\mathcal{I}^{\mu\nu}$ and momentum $P_{\mu}=\f{\hbar}{i}\f{\partial}{\partial Q^{\mu}}$ conjugate to nuclear coordinate $Q^{\mu}$.  Equation (\ref{eq:chi}) has the same form as the nuclear Schr\"odinger equation in the BO approximation \cite{moody1986} except the adiabatic potential energy surface and Mead-Truhlar vector potential \cite{mead1979} are replaced by their exact counterparts \cite{gidopoulos2014}
\begin{align}
\mathcal{E}(R) &= \langle \Phi_R | \hat{H}^{\rm BO} |  \Phi_R \rangle + \mathcal{E}_{\rm geo}(R)  
\label{eq:Epsilon} \\
A_{\mu}(R) &= \hbar\: \mathrm{Im} \langle \Phi_R |\partial_{\mu} \Phi_R\rangle {.}
\label{eq:A}
\end{align}
Here, $\partial_{\mu}=\f{\partial}{\partial Q^{\mu}}$ and $\mathcal{E}_{\rm geo}=\f{\hbar^2}{2} \mathcal{I}^{\mu\nu} g_{\mu\nu}$ is a geometric contribution to the potential energy surface \cite{requist2016}, which is analogous to a corresponding term in the BO approximation \cite{berry1989,berry1990}, and depends on the metric $g_{\mu\nu}$, the real part of the quantum geometric tensor \cite{berry1989}
\begin{align}
\mathcal{T}_{\mu\nu} = \langle \partial_{\mu} \Phi_R | (1-|\Phi_R\rangle \langle \Phi_R|) |\partial_{\nu} \Phi_R \rangle {.} \label{eq:qgt}
\end{align} 
The imaginary part is $1/\hbar$ times the Berry curvature $\mathcal{B}_{\mu\nu}$. 

The conditional electronic wavefunction acts like the BO wavefunction in standard DFT but includes all non\-adiabatic effects.  For example, the conditional electronic density can be calculated as \begin{align}
n_R(\mathbf{r}) = \langle \Phi_R | \sum_i \delta(\mathbf{r}-\mathbf{r}_i) | \Phi_R \rangle {.} 
\end{align}
From Eq.~(\ref{eq:chi}), we define the energy functional
\begin{align}
&E[n_R,\mathcal{T},\chi,A] = T_{\rm n,marg}[\chi,A] + \iint V_{\rm en} n_R(\mathbf{r}) |\chi|^2 d\mathbf{r} dR \nn \\
&\qquad\quad+\int \left( \mathcal{E}_{\rm geo}(R) + V_{\rm nn}(R) + F[n_R,\mathcal{T}] \right) |\chi|^2 dR {,} \label{eq:E:functional}
\end{align}
where $T_{\rm n,marg}$ is the ``marginal'' nuclear kinetic energy 
\begin{align}
T_{\rm n,marg} = \int \chi^*(R) \f{1}{2} \mathcal{I}^{\mu\nu} (P_{\mu}+A_{\mu})(P_{\nu}+A_{\nu}) \chi(R) dR \nn
\end{align}
and the constrained search procedure \cite{levy1979} is used to define the implicitly $R$-dependent functional
\begin{align}
F[n_R,\mathcal{T}] = \min_{\Psi \rightarrow (n_R,\mathcal{T})} \langle \Phi_R | \hat{T}_{\rm e} + \hat{V}_{\rm ee} | \Phi_R \rangle {.}
\label{eq:F}
\end{align}
We restrict ourselves to the bound states of isolated finite systems; external fields can be added straightforwardly.

\noindent \textbf{\textit{Theorem I}} --- The energy functional $E[n_R,\mathcal{T},\chi,A]$ is variational, i.e.~$E[n_R,\mathcal{T},\chi,A] \geq E_0$, and equality with the ground state energy $E_0$ is achieved for ground state $(n_R,\mathcal{T},\chi,A)$.  The domain of $E$ is the set of $(n_R,\mathcal{T},\chi,A)$ for which there exists a state $\Psi(r,R)$ with the correct particle exchange symmetry such that $\Psi\rightarrow (n_R,\mathcal{T},\chi,A)$ ($\Psi$-representability). \\
\textbf{\textit{Proof:}}
For any $\Psi$-representable $(\tilde{n}_{R},\tilde{\mathcal{T}})$, there exists a conditional wavefunction $\tilde{\Phi}_R$ which delivers the minimum in Eq.~(\ref{eq:F}) and for which $F[\tilde{n}_{R},\tilde{\mathcal{T}}] = \langle \tilde{\Phi}_R | \hat{T}_{\rm e}+\hat{V}_{\rm ee} | \tilde{\Phi}_R \rangle$.  Since for $\tilde{\Psi}=\tilde{\Phi}_R \tilde{\chi}$ we have the identity 
\begin{align}
T_{\rm n,marg}[\tilde{\chi},\tilde{A}] + \int \mathcal{E}_{\rm geo}(R) |\tilde{\chi}(R)|^2 dR = \langle \tilde{\Psi} | \hat{T}_n | \tilde{\Psi}\rangle  {,}
\end{align}
the right hand side of Eq.~(\ref{eq:E:functional}) is then equal to 
\begin{align}
\langle \tilde{\Psi} | \hat{T}_n + \hat{V}_{\rm en} + \hat{V}_{\rm nn} + \hat{T}_{\rm e} + \hat{V}_{\rm ee} | \tilde{\Psi} \rangle = \langle \tilde{\Psi} | \hat{H} | \tilde{\Psi} \rangle {.} 
\end{align}
Hence, the Rayleigh-Ritz variational principle implies
\begin{align}
E[\tilde{n}_{R},\tilde{\mathcal{T}},\tilde{\chi},\tilde{A}] = \langle \tilde{\Psi} | \hat{H} | \tilde{\Psi} \rangle \geq E_0 {.} \label{eq:RR}
\end{align}
To complete the proof, we need to show that the equality holds if $(\tilde{n}_{R},\tilde{\mathcal{T}},\tilde{\chi},\tilde{A})$ derive from a $\tilde{\Psi}$ which belongs to the ground state manifold.  By definition, the right-hand side of Eq.~(\ref{eq:F}) delivers the minimum of $\langle \Phi_R | \hat{T}_{\rm e} + \hat{V}_{\rm ee} | \Phi_R \rangle$ among states with $(\tilde{n}_{R},\tilde{\mathcal{T}})$ and so in particular
\begin{align}
\int F[\tilde{n}_{R},\tilde{\mathcal{T}}] |\tilde{\chi}(R)|^2 dR \leq \langle \Psi_0 | \hat{T}_{\rm e} + \hat{V}_{\rm ee} | \Psi_0 \rangle {,} \label{eq:F:inequality}
\end{align}
where $\Psi_0\!\rightarrow \!(n_{R0},\mathcal{T}_{0},\chi_0,A_0)$ is any state from the ground state manifold with $(n_{R0},\mathcal{T}_{0},|\chi_0|) = (\tilde{n}_{R},\tilde{\mathcal{T}},|\tilde{\chi}|)$.  Since
\begin{align*}
T_{\rm n,marg}[\tilde{\chi},\tilde{A}] + \int \left( \mathcal{E}_{\rm geo} + V_{\rm nn}\right. &+ \left.V_{\rm en} \tilde{n}_R(\mathbf{r}) d\mathbf{r} \right) |\tilde{\chi}(R)|^2 dR \nn \\
&= \langle \tilde{\Psi} | \hat{T}_{\rm n} + \hat{V}_{\rm nn} + \hat{V}_{\rm en} | \tilde{\Psi} \rangle \nn \\
&= \langle \Psi_0 | \hat{T}_{\rm n} + \hat{V}_{\rm nn} + \hat{V}_{\rm en} | \Psi_0 \rangle 
\end{align*}
if $(\tilde{n}_{R},\tilde{\chi},\tilde{A})=(n_{R0},\chi_0,A_0)$ to within a gauge transformation, then by adding the first and last members of the above chain of equalities to Eq.~(\ref{eq:F:inequality}), we obtain 
\begin{align}
E[\tilde{n}_{R},\tilde{\mathcal{T}},\tilde{\chi},\tilde{A}] \leq E_0 {,}
\end{align}
which together with Eq.~(\ref{eq:RR}) implies the desired result 
$E[\tilde{n}_{R},\tilde{\mathcal{T}},\tilde{\chi},\tilde{A}]=E_0$ for ground state $(\tilde{n}_{R},\tilde{\mathcal{T}},\tilde{\chi},\tilde{A})$. QED

The theorem is valid for degenerate and nondegenerate ground states, and it is an important point that the basic variables $(n_{R},\mathcal{T},\chi,A)$ may partially or completely resolve any degeneracy that is present, i.e.~it is generally the case that not all of the states in a degenerate ground state manifold have the same $(n_{R},\mathcal{T},\chi,A)$.  An example occurs in the model triatomic  molecule studied in Ref.~\onlinecite{requist2016}, where  $n_{R}$  and $\mathcal{T}_{\mu\nu}$ single out a unique degenerate ground state.  As in DFT, we now need a workable procedure for minimizing the functional $E[n_{R},\mathcal{T},\chi,A]$. \\ 
\noindent \textbf{\textit{Minimization scheme.}}  The ground state $(n_{R},\mathcal{T},\chi,A)$ can be calculated by solving the following three sets of equations self consistently.  

(i) The conditional Kohn-Sham equations 
\begin{align}
\LB \f{\mathbf{p}^2}{2m} + v_{\rm en}(\mathbf{r},R) + v_{\rm hxc}(\mathbf{r},R) \RB \psi_{Rk\s}(\mathbf{r}) = \epsilon_{Rk\s} \psi_{Rk\s}(\mathbf{r}) 
\label{eq:KS}
\end{align}
determine $n_R(\mathbf{r})= \sum_{k\s} f_{Rk\s} |\psi_{Rk\s}(\mathbf{r})|^2$, where $f_{Rk\s}$ is the occupation number of the state $\psi_{Rk\s}$. The definition of the Hartree-exchange-correlation (hxc) potential is 
\begin{align}
v_{\rm hxc}(\mathbf{r},R) = \f{\delta E_{\rm hxc}}{\delta n_R(\mathbf{r})}  {,} \label{eq:vhxc}
\end{align}
where $E_{\rm hxc}[n_R,\mathcal{T}] = F [n_R,\mathcal{T}] - T_{\rm s}[n_R]$ is the conditional hxc energy and the conditional kinetic energy functional of noninteracting electrons is defined as
\begin{align}
T_{\rm s}[n_R] &= \min_{\Phi_{R\rm s}\rightarrow n_R} \langle \Phi_{R\rm s} | \hat{T}_{\rm e} | \Phi_{R\rm s} \rangle {.}
\label{eq:Ts}
\end{align}
Here, the search is over Slater determinants $\Phi_{R\rm s}(r)$ (or over ensembles of degenerate Slater determinants if fractional $f_{Rk\sigma}$ are needed \cite{valone1980,dreizler1990}).  The stationary condition $\delta E/\delta n(\mathbf{r},R)=0$ subject to the fixed electron number constraint $\int |\chi(R)|^2 \delta n_R(\mathbf{r}) d\mathbf{r} dR = 0$ gives
\begin{align}
\left\{v_{\rm en}(\mathbf{r},R) + v_{\rm hxc}(\mathbf{r},R) + \f{\delta T_{\rm s}}{\delta n(\mathbf{r},R)} \right\} |\chi(R)|^2 &= 0 {.}
\end{align}
For all $R$ for which $|\chi(R)|\neq 0$, this is exactly the stationary condition of standard DFT for noninteracting electrons in a potential $v_R(\mathbf{r})=v_{\rm en}(\mathbf{r},R) +v_{\rm hxc}(\mathbf{r},R)$, which implies that $n_R(\mathbf{r})$ can be calculated by solving Eq.~(\ref{eq:KS}).  Since $v_{\rm hxc}(\mathbf{r},R)$ is defined analogously to $v_{\rm hxc}(\mathbf{r})$ in DFT, the only difference being the extra $\mathcal{T}_{\mu\nu}$ dependence, we expect the potentials to be similar for regions of $R$ where nonadiabatic effects are small.  This was the motivation for choosing $\mathcal{T}_{\mu\nu}$ as a basic variable and for deriving conditional Kohn-Sham equations in the form of Eq.~(\ref{eq:KS}), which does not correspond to the limit $V_{ee}\rightarrow 0$ in Eq.~(\ref{eq:Phi}).  As a first approximation for $E_{\rm hxc}$, we can simply substitute $n_R(\mathbf{r})$ in place of $n(\mathbf{r})$ in existing DFT functionals.  The optimized effective potential equation in Ref.~\onlinecite{gidopoulos2014} also provides a way to approximate $v_{\rm en}(\mathbf{r},R) +v_{\rm hxc}(\mathbf{r},R)$. 

(ii) The stationary condition with respect to variations of $\chi(R)$ yields the nuclear Schr\"odinger equation, Eq.~(\ref{eq:chi}).

(iii) The ground state quantum geometric tensor $\mathcal{T}_{\mu\nu}$ could be determined by direct minimization; however, we find it more useful in practice to calculate $\mathcal{T}_{\mu\nu}$ indirectly from a set of $N$ auxiliary functions $\lambda^{\mu}(R)$, where $N$ is the dimension of the nuclear configuration space $\mathcal{Q}$.  This means we consider $E$ as a functional of $(\lambda^{\mu},\partial_{\nu}\lambda^{\mu})$ instead of $\mathcal{T}_{\mu\nu}$ by a straightforward generalization of theorem I.  The $\lambda^{\mu}$ then satisfy the Euler-Lagrange equations
\begin{align}
\f{\delta E}{\delta \lambda^{\mu}} - \f{d}{dQ^{\nu}} \f{\delta E}{\delta (\partial_{\nu} \lambda^{\mu})} = 0 {.}
\label{eq:EL}
\end{align}
To calculate $\mathcal{T}_{\mu\nu}$ from $\lambda^{\mu}$, we start by defining a different quantum geometric tensor $T_{\mu\nu}=\langle \partial_{\mu} \Phi | (1-|\Phi\rangle \langle \Phi|) |\partial_{\nu} \Phi \rangle$, where the derivatives $\partial_{\mu}$ are taken with respect to canonical coordinates $\xi^{\mu}=(q^1,\ldots, q^n|p_1,\ldots, p_n)$ for the projective Hilbert space $\mathcal{P}_{\Phi}$ of the electronic function $|\Phi\rangle$ \cite{berry1989,provost1980}, which, for convenience, and in accordance with the way calculations are done, has been represented by a finite basis with dimension $2n\geq N$.  The map $\mathcal{Q} \xrightarrow{\Phi_R} \mathcal{P}_{\Phi}$ defines functions $\xi^{\mu}=\xi^{\mu}(R)$.  We now assume that the functions $\lambda^{\mu}$ are related to the $\xi^{\mu}$ by a local coordinate transformation $\xi^{\mu} \rightarrow \lambda^{\alpha} = (\lambda^1, \ldots, \lambda^{N}, \lambda^{N+1}, \ldots, \lambda^{2n})$ such that $\lambda^{\alpha}(\xi^{\mu}(R))$ obey the conditions
\begin{align}
\f{\partial \lambda^{\alpha}}{\partial Q^{\mu}} = 0 \quad \textrm{for all} \; \mu, R\; \textrm{and}\; \alpha> N   {.}
\end{align} 
The first $N$ functions $\lambda^{\mu}(R)$ then determine $\mathcal{T}_{\mu\nu}(R)$ by 
\begin{align}
\mathcal{T}_{\mu\nu} = T_{\alpha\beta} \f{\partial \lambda^{\alpha}}{\partial Q^{\mu}} \f{\partial \lambda^{\beta}}{\partial Q^{\nu}} {,} \label{eq:pullback:g}
\end{align}
where $T_{\alpha\beta} = \langle \partial_{\lambda^{\alpha}} \Phi | \big( 1-|\Phi\rangle \langle \Phi | \big) | \partial_{\lambda^{\beta}} \Phi \rangle$.
Lastly, we note that we can calculate the $N_{\rm n}$-body nuclear current as
\begin{align}
J^{\mu}(R) 
&= \mathcal{I}^{\mu\nu} \LB \hbar\, \mathrm{Im}\chi^*\partial_{\nu} \chi + A_{\nu}|\chi|^2 \RB {.} \label{eq:J}
\end{align}

This scheme provides a way to calculate the ground state $(n_R,\mathcal{T}_{\mu\nu},\chi,A_{\mu})$.  Equations (i-iii) implicitly couple many-body electronic structure to induced electromagnetism in the nuclear Schr\"odinger equation.

\noindent \textbf{\textit{Example calculation}:} The $E\otimes e$ Jahn-Teller model consists of a doubly degenerate electronic level coupled to two degenerate vibrational normal modes, whose amplitudes are conventionally denoted $Q_2$ and $Q_3$.  The Hamiltonian is $\hat{H} = (\hbar^2/2\mathcal{M})(P_2^2+P_3^2) + (\mathcal{K}/2) (Q_2^2+Q_3^2) + \hat{H}_{\rm en}$ with the electronic-vibrational coupling given by
\begin{align}
H_{\rm en} = g\LB \bar{cc} Q_2 & - Q_3 \\ -Q_3 & -Q_2 \ear \RB \label{eq:Hen}
\end{align}
in the basis $(|o\rangle,|e\rangle)$ of odd/even states.  Transforming to coordinates $(Q,\eta)=(\sqrt{Q_2^2+Q_3^3},\tan^{-1}(Q_3/Q_2))$, and applying the unitary matrix $U=\f{1}{\sqrt{2}} (( i , -i ), ( 1 , 1 ))$ gives 
\begin{align}
U^{\dag} H_{\rm en} U =  g\LB \bar{cc} 0 & - Q e^{-i\eta} \\ - Q e^{i\eta} & 0 \ear \RB {,} 
\end{align}
in the basis $|\pm\rangle = \f{1}{\sqrt{2}} (|e\rangle\pm i |o\rangle)$.  The adiabatic potential energy surface $\mathcal{E}^{\rm BO} = \f{\mathcal{K}}{2} (Q-\f{g}{\mathcal{K}})^2 - \f{g^2}{2\mathcal{K}}$ is the well-known sombrero potential with a conical intersection at $Q=0$ and classical Jahn-Teller stabilization energy $\f{g^2}{2\mathcal{K}}$.

The ground state manifold is spanned by the two states with angular momentum quantum numbers $j=\pm \f{1}{2}$ (see Supplemental Material).  
For definiteness, we take the $j=\f{1}{2}$ state, which can be written $|\Psi\rangle = a(Q) |+\rangle + b(Q) e^{i\eta}|-\rangle$.  Defining $\chi=\sqrt{a^2+b^2}$ and $\theta=2\tan^{-1}(b/a)$ gives $|\Phi_R\rangle = \cos\f{\theta}{2} |+\rangle + \sin\f{\theta}{2} e^{i\varphi} |-\rangle$ with $\varphi=\eta$.

To calculate the ground state $(n_R,\mathcal{T},\chi,A)$, we need to self-consistently solve equations (i-iii).  Since the sum of the occupations of $|e\rangle$ and $|o\rangle$ orbitals is 1, we choose $n_R \equiv n_{Re}-n_{Ro} = \sin\theta \cos\eta$ as the single independent density variable.  In the adiabatic limit $\theta\rightarrow\f{\pi}{2}$, this gives $\cos\eta$, which means e.g.~that the even orbital is fully occupied if the nuclei are distorted along normal mode $Q_2$, cf.~Eq.~(\ref{eq:Hen}).  Since $n_R$ is completely determined by $\theta$ and $\eta$, and $\theta$ will be determined by equation (iii), it is not actually necessary here to set up the conditional Kohn-Sham equations, and we can proceed to equation (ii) for $\chi$, which from Eq.~(\ref{eq:chi}) is found to be
\begin{align}
-\f{\hbar^2}{2\mathcal{M}} \left[ \f{1}{Q} \f{d}{dQ} Q \f{d}{dQ} - \f{1}{Q^2} \sin^4\f{\theta}{2} \right] \chi + \mathcal{E} \chi = E \chi {,} \label{eq:chi:explicit}
\end{align}
where 
\begin{align}
\mathcal{E} &= \f{\mathcal{K}}{2} Q^2 - g Q\sin\theta + \mathcal{E}_{\rm geo} {,} \nn \\
\mathcal{E}_{\rm geo} &= \f{\hbar^2}{2\mathcal{M}} \left[ \f{1}{4} \LB \f{d\theta}{dQ} \RB^2 + \f{\sin^2\theta}{4Q^2} \right] {,} \label{eq:energysurface}
\end{align}
and we used $\mathcal{I}^{QQ} = \f{1}{\mathcal{M}}$, $\mathcal{I}^{\eta\eta}=\f{1}{\mathcal{M}Q^2}$, $A_{\eta} = \hbar \sin^2\f{\theta}{2}$ and $g=\mathrm{Re}\mathcal{T}$ from $\mathcal{T}$ in the $(\eta,Q)$ basis 
\begin{align}
\mathcal{T} = \f{1}{4} \LB \bar{cc} \sin^2\theta & -i \sin\theta \,\partial_Q\theta \\ i  \sin\theta \,\partial_Q\theta & (\partial_Q\theta)^2\ear \RB {.} \label{eq:T}
\end{align}
For convenience, we use $(\varphi,\theta)$ coordinates, which are related to the canonical coordinates $q=\varphi$ and $p=\hbar \sin^2\f{\theta}{2}$.
The crucial difference between the exact potential energy surface $\mathcal{E}$ and the BO potential energy surface $\mathcal{E}^{\rm BO}$ is the appearance of the factor $\sin\theta$ multiplying the term $-gQ$ responsible for the Jahn-Teller distortion in the static picture.  Since $\theta(Q)$ deviates from its constant adiabatic value $\f{\pi}{2}$ due to nonadiabatic mixing between BO states, the $\sin\theta$ factor weakens the electronic-vibrational coupling and the Jahn-Teller stabilization energy with respect to its classical adiabatic value $\f{g^2}{2\mathcal{K}}$.  

From the imaginary part of $\mathcal{T}$, we can obtain the Berry curvature $\mathcal{B}_{Q\eta}=\f{\hbar}{4} \sin\theta \partial_Q \theta$ and calculate the molecular geometric phase on a circular path of radius $Q$ bounding the disk $\mathcal{S}$ as \cite{requist2016,gidopoulos2014,min2014,berry1984}
\begin{align}
\gamma(Q) &= \f{1}{\hbar} \int_{\mathcal{S}} \mathcal{B}_{\mu\nu} dQ^{\mu}\wedge dQ^{\nu} = \pi \left[1-\cos\theta(Q)\right] {,}
\label{eq:gamma}
\end{align}
which coincides with the geometric phase of a pseudospin precessing with  polar angle $\theta=\theta(Q)$ on the Bloch sphere.

\begin{figure}[tb!]
\includegraphics[width=0.95\columnwidth]{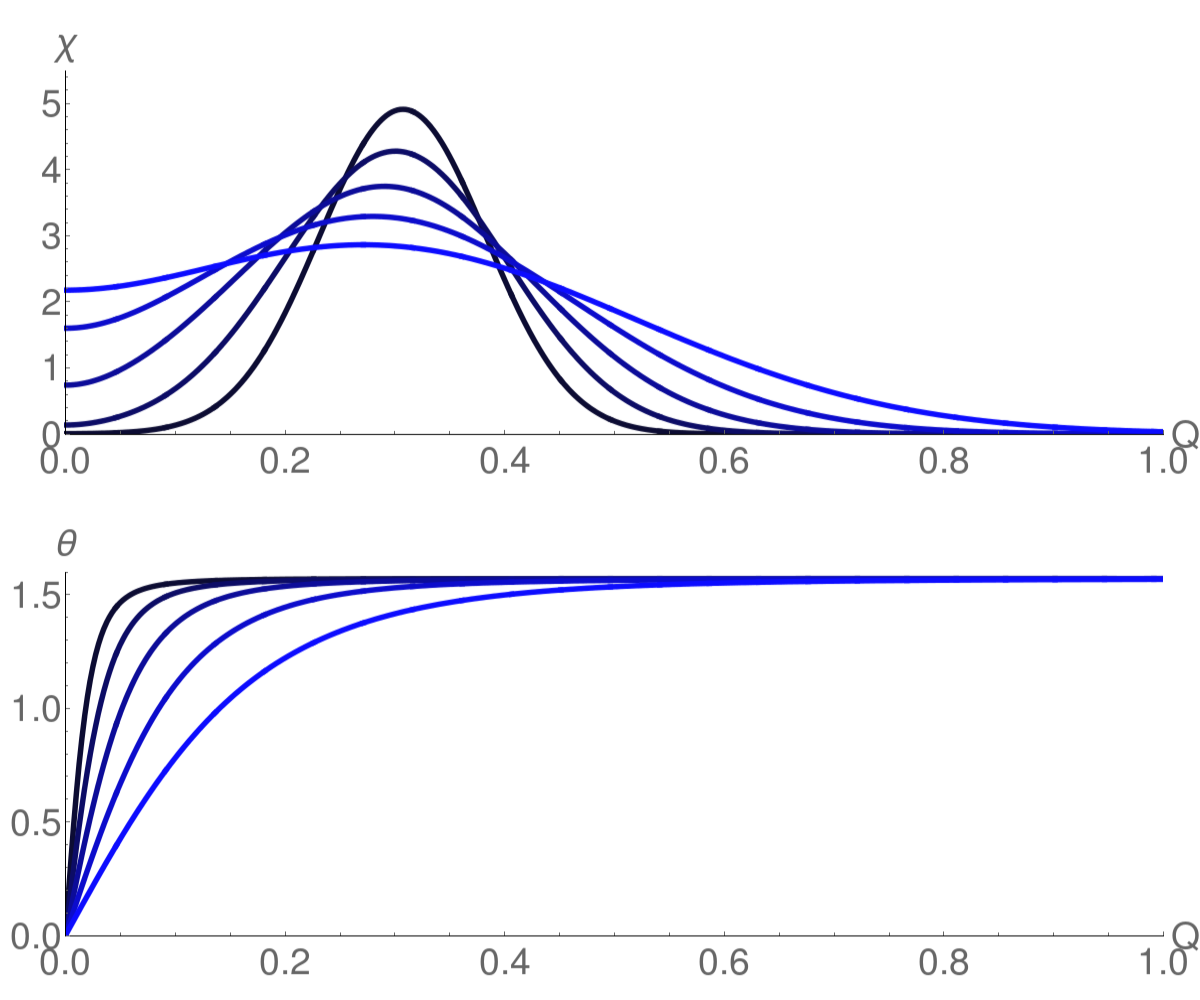}
\caption{The nuclear wavefunction $\chi(Q)$ and electronic variable $\theta(Q)$ for $g=10^{-1/2}$~eV/\AA, $\mathcal{K}=1$eV/\AA$^2$ and the following values of $\mathcal{M}$: $(10^{-1/2},10^{-1},10^{-3/2},10^{-2},10^{-5/2})\hbar^2$eV/\AA$^2$ (color code: light to dark blue).  Distances are in \AA ngstroms.}
\label{fig:chitheta}
\end{figure}

Turning to (iii), we note that the two $\lambda^{\mu}$ variables are $\lambda^1=q$ and $\lambda^2=p$.  $\lambda^1=\eta$ is already known, and since $\theta$ uniquely determines 
$\lambda^2=\hbar \sin^2\f{\theta}{2}$, we can derive a single Euler-Lagrange equation for $\theta$ instead of $\lambda^2$:  
\begin{align}
Q^2 \f{d^2\theta}{dQ^2} + \LB 1+ Q \f{d}{dQ}\log|\chi|^2 \RB Q \f{d\theta}{dQ}- \sin\theta &\nn\\
+ \f{4g\mathcal{M}}{\hbar^2} Q^3 \cos\theta &= 0 {.}
\label{eq:theta}
\end{align}
The original linear system has thus been transformed to a pair of coupled nonlinear differential equations, Eqs.~(\ref{eq:chi:explicit}) and (\ref{eq:theta}), which are to be solved with the boundary conditions $\chi(\infty)=0$, $\theta(0)=0$ and $\theta(\infty)=\f{\pi}{2}$.  The latter condition is directly related to a topological invariant of the Berry curvature, namely, the surface integral of $\mathcal{B}_{Q\eta}$ over the $(Q_2,Q_3)$ plane must be a multiple of $\pi$ because the plane can be compactified to a sphere and $\mathcal{B}_{Q\eta}$ vanishes as $Q\rightarrow\infty$.  Therefore, in accordance with Eq.~(\ref{eq:gamma}), $\theta(\infty)=\f{\pi}{2}$ gives the $j=\f{1}{2}$ state. The numerical solution of Eqs.~(\ref{eq:chi:explicit})  and (\ref{eq:theta}) is shown for several values of $\mathcal{M}$ in Fig.~\ref{fig:chitheta}.  We have also calculated $\chi(Q)$ and $\theta(Q)$ after solving the model by exact diagonalization \cite{longuet-higgins1958}, verifying that the same results are obtained by both methods.  

In the BO limit $\mathcal{M}\rightarrow\infty$, $\theta(Q)$ jumps discontinuously from 0 to $\pi/2$ (see Fig.~\ref{fig:chitheta}).  This is the well-known result that in the BO approximation the Berry curvature is a Dirac delta function $\f{h}{2} \delta(Q_2) \delta(Q_3)$, which can be neatly attributed to the flux of an infinitesimal Aharonov-Bohm flux tube located at the conical intersection. In contrast, the smooth rise of $\theta$ from 0 to $\pi/2$ in the exact calculation is a result of the ``smearing out'' of the Aharonov-Bohm flux tube due to nonadiabatic effects \cite{requist2016}.  

The above approach can be used to identify the interactions that control the smearing width.  The electronic-vibrational interaction energy $-\iint g Q \sin\theta |\chi|^2 \,QdQd\eta$ favors a peaked Berry curvature, and if it were the only relevant term, minimizing the energy with respect to~$\mathcal{B}_{Q\eta}(Q)$ would yield a delta function.  The geometric term $\int \mathcal{E}_{\rm geo} |\chi|^2 \,QdQd\eta$ and the centrifugal repulsion $\f{\hbar^2}{2\mathcal{M}} \iint \f{1}{Q^2} \sin^4\f{\theta}{2} |\chi|^2 \,QdQd\eta$, which both originate from the nuclear kinetic energy, favor a broader profile.  Hence, the true profile of the Berry curvature results from a compromise between Jahn-Teller stabilization energy and kinetic repulsion.

The conditional density is isotropic in $\eta$ at the origin $(n_{Re}=n_{Ro})$ but becomes anisotropic for $Q>0$.  This response is weakened by nonadiabatic effects embodied in the $\sin\theta$ factor of $n_R$.  The size of the region where the anisotropic response is significantly weakened correlates with the smearing width of the Berry curvature.  The identity $\langle \Phi_R | \hat{J}_z | \Phi_R \rangle = L_z(Q)+l_z(Q)=\f{1}{2}$ with $L_z(Q)=\sin^2\f{\theta}{2}$ (see Supplemental Material) implies that the nuclei transfer angular momentum to the electrons as $Q\rightarrow 0$ and that the electronic state must cross over from the anisotropic adiabatic state $|\Phi_R^{\rm BO}\rangle = \f{1}{\sqrt{2}} (|+\rangle + e^{i\eta}|-\rangle)$ to the isotropic current-carrying state $|+\rangle$ at $Q=0$, resulting in a weakened density response near the origin.  The width of the crossover region is given by the characteristic scale of the rise of $\theta$ (see Fig.~\ref{fig:chitheta}) and is therefore determined by the same nonadiabatic effects as the Berry curvature smearing width.  The exact conditional density is smooth in contrast to the adiabatic case, where it is nonanalytic (``topologically scarred'') at the conical intersection \cite{baer2010}.

The theory presented here couples electronic density-functional theory to the nuclear Schr\"odinger equation in a rigorously exact way.  If the full solution of the nuclear Schr\"odinger equation is prohibitive, approximations such as the trajectory-based methods developed within the exact factorization scheme \cite{abedi2014,agostini2015b} can be used to solve the nuclear part of the problem.  Exact factorization-based DFT can be used to include nonadiabatic quantum nuclear effects in systems with many electrons and nuclei, such as  large molecules, models of water and ice, and nanostructures, if accurate functional approximations can be found for $E_{\rm hxc}[n_R,\mathcal{T}]$.  One can  hope that the small parameter $m_{\rm e}/m_{\rm n}$, the ratio of electronic and nuclear masses, can be used to derive asymptotic approximations for the $\mathcal{T}$-dependence.  While quantum nuclear effects are small in the ground states of most systems, they are utterly inescapable in many real-time physical and chemical processes, which fall within the scope of the time dependent version of the theory presented here.

\bibliography{bibliography}

\end{document}